\documentclass[aps,pra,twocolumn,superscriptaddress]{revtex4-2}
\usepackage{color}
\usepackage{array}
\usepackage{textcomp}\usepackage{nccmath}
\usepackage{bm}
\usepackage{braket}
\usepackage{amsmath}
\usepackage{amssymb}
\usepackage{upgreek}
\usepackage{multirow}
\usepackage{makecell}
\usepackage{graphicx}
\usepackage{hyperref}
\hypersetup{colorlinks=true, citecolor=blue, urlcolor=blue, linkcolor=blue}

\begin{document}

\title{Enhanced Persistent Orientation of Asymmetric-Top Molecules Induced by Cross-Polarized Terahertz Pulses}
\author{Long Xu}
\email{xulong@xmu.edu.cn}
\affiliation{Department of Physics, Xiamen University, Xiamen 361005, China}
\affiliation{AMOS and Department of Chemical and Biological Physics, The Weizmann Institute of Science, Rehovot 7610001, Israel}
\author{Ilia Tutunnikov}
\email{iliat@mit.edu}
\affiliation{AMOS and Department of Chemical and Biological Physics, The Weizmann Institute of Science, Rehovot 7610001, Israel}
\affiliation{Department of Chemistry, Massachusetts Institute of Technology, Cambridge, Massachusetts 02139, USA}
\author{Yehiam Prior}
\email{yehiam.prior@weizmann.ac.il}
\affiliation{AMOS and Department of Chemical and Biological Physics, The Weizmann Institute of Science, Rehovot 7610001, Israel}
\author{Ilya Sh. Averbukh}
\email{ilya.averbukh@weizmann.ac.il}
\affiliation{AMOS and Department of Chemical and Biological Physics, The Weizmann Institute of Science, Rehovot 7610001, Israel}
\affiliation{Department of Chemistry, University of British Columbia, Vancouver, British Columbia V6T 1Z1, Canada}

\begin{abstract}
We investigate the persistent orientation of asymmetric-top molecules induced by time-delayed THz pulses that are either collinearly or cross polarized. Our theoretical and numerical results demonstrate that the orthogonal configuration outperforms the collinear one, and a significant degree of persistent orientation - approximately 10\% at 5\,K and nearly 3\% at room temperature - may be achieved through parameter optimization. The dependence of the persistent orientation factor on temperature and field parameters is studied in detail. The proposed application of two orthogonally polarized THz pulses is both practical and efficient. Its applicability under standard laboratory conditions lays a solid foundation for future experimental realization of THz-induced persistent molecular orientation.
\end{abstract}

\maketitle

\section{Introduction}
Molecular alignment and orientation induced by short laser pulses have attracted widespread interest across various disciplines within ultrafast science. This is mainly due to their pivotal role in numerous physical processes, such as laser-induced ionization, high-order harmonic generation, molecular imaging, etc. Existing methods for molecular rotational control and their applications have been reviewed extensively over the past two decades \cite{Stapelfeldt2003Colloquium,Ohshima2010Coherent,Fleischer2012Molecular,Lemeshko2013Manipulation,Koch2019Quantum,Lin2020Review}.
Besides molecular alignment, various strategies have been proposed and implemented to achieve molecular orientation. 
One such tool employs optical-frequency two-color laser pulses consisting of a fundamental wave and its second harmonic. Such pulses were shown to be effective in inducing molecular orientation via two distinct intensity-dependent mechanisms \cite{Vrakking1997Coherent,Dion1999Two,Kanai2001Numerical,Takemoto2008Fixing,De2009Field,Oda2010All,Wu2010Field,Zhang2011Controlling,Frumker2012Oriented,Spanner2012Mechanisms,Znakovskaya2014Transition,lin2018All,Mun2018Improving,Mun2019Orientation,Mun2020Field,Mellado2020Linear,Shuo2020Optimal,Xu2021longlasting,Xu2021Three,Hossain2022Stronger,Xu2022Ionization}.
At low (nonionizing) intensities, the dominant mechanism depends on the interaction of the two-color pulse with the molecular hyperpolarizability, inducing asymmetric torques that orient the molecules \cite{Vrakking1997Coherent,Dion1999Two,Kanai2001Numerical,Takemoto2008Fixing,De2009Field,Oda2010All,Wu2010Field,Zhang2011Controlling,Spanner2012Mechanisms,Znakovskaya2014Transition,lin2018All,Mun2018Improving,Mun2019Orientation,Mun2020Field,Mellado2020Linear,Shuo2020Optimal,Xu2021longlasting,Xu2021Three,Hossain2022Stronger,Xu2022Ionization}.
The other mechanism, which is dominant at high (ionizing) intensities, is caused by the different probability of ionization along or against the polarization direction of the asymmetric electric field generated by the collinearly polarized two-color pulse \cite{Frumker2012Oriented,Spanner2012Mechanisms,Znakovskaya2014Transition,Xu2022Ionization}.
Another method combines weak electrostatic and intense non-resonant laser fields \cite{Friedrich1999Enhanced,Sakai2003Controlling,Goban2008Laser,Ghafur2009Impulsive,Holmegaard2009Laser,Mun2014Laser,Takei2016Laser,Omiste2016Theoretical,Thesing2017Time}. 
Advances in Terahertz (THz) technology paved the way for achieving molecular orientation using pulsed THz fields, either standalone \cite{Harde1991THz,Averbukh2001Angular,Machholm2001Field,Fleischer2011Molecular,Kitano2013Orientation, Damari2016Rotational,Babilotte2016Observation,Xu2020longlasting} or in conjunction with optical-frequency pulses \cite{Daems2005Efficient,Gershnabel2006Orientation,Egodapitiya2014Terahertz,Tutunnikov2022echo}.
Recently, attention has also shifted towards the study of enantioselective orientation, which employs laser or THz pulses with twisted polarization \cite{Yachmenev2016Detecting,Gershnabel2018Orienting,Tutunnikov2018Selective,Milner2019Controlled,Tutunnikov2019Laser,Tutunnikov2020Observation,Tutunnikov2021Enantioselective,Xu2022enantioselective}, as well as two-color cross-polarized laser pulses \cite{Takemoto2008Fixing,Xu2021Three}. 

In addition to the well-known transient orientation that appears immediately following excitation by the laser pulses, a new phenomenon of \textit{persistent orientation} has recently been studied. The effect was observed for the first time in chiral molecules excited by laser pulses with twisted-polarization  \cite{Tutunnikov2019Laser,Tutunnikov2020Observation}. Furthermore, it has been predicted in non-linear molecules stimulated by THz pulses \cite{Xu2020longlasting,Tutunnikov2021Enantioselective} or two-color laser pulses \cite{Xu2021longlasting,Xu2021Three,Xu2022Ionization}. In this context, ``persistent'' refers to a nonzero average orientation that persists on a time scale several orders of magnitude longer than the duration of the applied external field.

While persistent orientation in chiral molecules has been experimentally achieved using polarization-shaped laser pulses \cite{Tutunnikov2020Observation}, similar effects are yet to be realized using THz or two-color laser pulses. One reason is that the value of the persistent orientation is generally small, especially at high temperatures. Thus, an efficient approach is required to enhance the effect and make it experimentally feasible.

Here, we investigate the persistent orientation of asymmetric-top molecules excited by THz pulses. We show that the persistent orientation may be significantly improved using a pair of time-delayed cross-polarized THz pulses, compared to excitation by collinearly polarized pulses. We further investigate the dependencies on the field amplitude, the polarization angle, the time delay between pulses, and the temperature. 
The paper is organized as follows. Section \ref{sec:methods} introduces the theoretical methods. The persistent orientation induced by a single THz pulse and a pair of time-delayed THz pulses are presented in
Sec. \ref{sec:single_pulse} and Sec. \ref{sec:delayed_pulse}, respectively. Finally, we summarize the work in Sec. \ref{sec:conclusions}.

\section{Theoretical methods}\label{sec:methods}

Asymmetric-top molecules are modeled as rigid bodies with three orthogonal inertia axes denoted by $a$, $b$, and $c$, with the moments of inertia $I_{a}<I_{b}<I_{c}$. We perform quantum mechanical and classical simulations of the molecular rotations following the THz excitation. 
The model Hamiltonian is $H(t)=H_{r}+H_{\mathrm{int}}(t)$ \cite{Krems2018Molecules,Koch2019Quantum}, where $H_{r}$ is the rotational kinetic energy Hamiltonian and $H_{\mathrm{int}}(t)=-\bm{\mu}\cdot\mathbf{E}(t)$ describes the field-molecule interaction. 
Here $\bm{\mu}$ is the molecular dipole moment vector, and $\mathbf{E}(t)$ is the electric field vector.
The symmetric-top eigenfunctions $|JKM\rangle$ are used as a basis set \cite{zare1988Angular}.
The quantum numbers $J$, $K$, and $M$ correspond to the total angular momentum and its projections on the molecular $a$ and the laboratory $X$ axes, respectively.
The nonzero matrix elements of $H_{r}$ are given by \cite{zare1988Angular}
\begin{flalign}
&\langle JKM|H_{r}|JKM\rangle=\frac{B+C}{2} \left[J(J+1)-K^{2}\right]+AK^{2},\nonumber\\
&\langle JKM|H_{r}|J,K\pm 2,M\rangle  =\frac{B-C}{4}f(J,K\pm 1),\label{eq:HR}
\end{flalign}
where $A=\hbar^{2}/2I_{a}$, $B=\hbar^{2}/2I_{b}$, $C=\hbar^{2}/2I_{c}$, and $f(J,K)=\sqrt{(J^{2}-K^{2})[(J+1)^{2}-K^{2}]}$. The eigenfunctions of the asymmetric-top molecule, $|J\kappa M\rangle=\sum_{K}c_{K}^{(J\kappa M)}|JKM\rangle$, with $-J\leq\kappa\leq J$, can be obtained by numerical diagonalization of $H_{r}$.

The finite initial rotational temperature is accounted for using the random phase wave functions method \cite{Kallush2015Orientation}. In this method, the initial state is defined as
\begin{align}\label{eq:initial_state}
|\Psi_{n}(t_i)\rangle=\sum_{J\kappa M}\sqrt{\frac{e^{-\varepsilon_{J\kappa M}/(k_{B}T)}}{\mathcal{Z}}}e^{i\varphi_{n,J\kappa M}}|J\kappa M\rangle,
\end{align}
where $t_i$ is the initial time, $\varepsilon_{J\kappa M}$ is the rotational kinetic energy of $|J\kappa M\rangle$, $k_{B}$ is the Boltzmann constant, $T$ is the temperature,  and $\mathcal{Z}$ is the partition sum. 
Here $\varphi_{n, J\kappa M}\in[0,2\pi)$ is a random number and the sum runs over all the thermally populated eigenstates $|J\kappa M\rangle$.
The time-dependent Schr\"{o}dinger equation $i\hbar \partial_t |\Psi_{n}(t)\rangle = H(t)|\Psi_{n}(t)\rangle$ is solved using numerical exponentiation of the Hamiltonian matrix (see Expokit \cite{sidje1998Expokit}).
The expectation value gives the degree of orientation of the molecular $a$ axis,
\begin{align}
\braket{\cos(\theta_{aj})}(t)
=\frac{1}{N}\sum\limits _{n=1}^{N}\langle \Psi_{n}(t)|\bm{a}\cdot \mathbf{e}_j|\Psi_{n}(t)\rangle,\label{eq:Quantum-Boltzmann-distribution-1}
\end{align}
where $N$ is the number of initial states, $\mathbf{e}_{j}$ is a unit vector along the laboratory $j$ ($j=X,Y,Z$) axis, $\bm{a}$ is a unit vector along the molecular $a$ axis, and $\theta_{aj}$ is the angle between the molecular $a$ axis and the laboratory $j$ axis. Given the temperatures considered ($T \geq 5\,$K), we found that $N=100$ is sufficient for convergent results, where the absolute error in the degree
of orientation is less than 0.001.

Here we use formaldehyde ($\mathrm{CH_2O}$) as an example asymmetric-top molecule. In this molecule, the molecular dipole moment points along the molecular $a$ axis.
Table \ref{tab:Molecular-properties} summarizes the molecular properties as taken from NIST [density functional theory (DFT), Coulomb-attenuating method with Becke three-parameter Lee-Yang-Parr functional and augmented correlation-consistent polarized valence triple zeta Gaussian basis set (CAM-B3LYP/aug-cc-pVTZ)] \cite{johnson2019nist}.
Since formaldehyde is a planar molecule belonging to the $C_{2v}$ point symmetry group, the statistical factor due to nuclear spin \cite{Bechtel2005Nuclear} should be included in the sum in Eq. \eqref{eq:initial_state}.
According to Eq. \eqref{eq:HR}, the rotational states with odd and even $K$ are decoupled.
As a result, the initial rotational $\ket{J\kappa M}$ states formed by $\ket{JKM}$ states with odd $K$ have triple the statistical weight of $\ket{J\kappa M}$ states formed by $\ket{JKM}$ with even $K$ \cite{Bechtel2005Nuclear}.

\begin{table}[!t]
\setlength{\abovecaptionskip}{0.cm}
\centering
\caption{\bf Molecular properties of $\mathrm{CH_2O}$: moments of inertia (in atomic units) and nonzero elements of the dipole moment (in Debye) in the frame of principal axes of inertia.
\label{tab:Molecular-properties}}
\begin{centering}\setlength{\tabcolsep}{2mm}{
\begin{tabular}{ccc}\hline
 Molecule                  & Moments of inertia            & Dipole components                                    \\ \hline
\multirow{3}{2cm}{\makecell{Formaldehyde \\$\mathrm{CH_2O}$}} 
& $I_{a}=11560$ & $\mu_{a}=2.389$ 
   \\
                    &   $I_{b}=84122$                                          &                   \\
                 &      $I_{c}=95682$                                        &                                \\\hline
\end{tabular}}
\end{centering}
\end{table}

In the classical limit, we model the ensemble behavior using the Monte Carlo approach.
Initially, sample molecules are isotropically distributed in space, their angular velocities ($\Omega_{i}$) are random and follow the Boltzmann distribution $P(\Omega_{i})\propto\exp[-I_{i}\Omega_{i}^{2}/(2k_{B}T)]$, where $i=a,b,c$.
The rotational dynamics of each sample molecule is determined by Euler's equations in the rotating molecular frame \cite{Goldstein2002Classical}
\begin{equation}
\mathbf{I}\bm{\dot{\Omega}}=(\mathbf{I}\bm{\Omega})\times\bm{\Omega}+\mathbf{T},\label{eq:Eulers-equations}
\end{equation}
where $\mathbf{I}=\mathrm{diag}(I_{a},I_{b},I_{c})$ is the moment
of inertia tensor, $\bm{\Omega}=(\Omega_{a},\Omega_{b},\Omega_{c})$
is the angular velocity vector, and $\mathbf{T}=(T_{a},T_{b},T_{c})$ is
the torque due to the interactions with the THz field, given by $\mathbf{T}=\bm{\mu}\times \mathbf{E}_\mathrm{mol}(t)$.
Here, $\mathbf{E}_\mathrm{mol}(t)$ is the electric field vector expressed in the frame of principal axes of inertia. 
A quaternion,
$q=(q_{0},q_{1},q_{2},q_{3})$ parametrizes the relationship between the molecular and laboratory frames \cite{Kuipers1999Quaternions,Coutsias2004The}.
The equation of motion for a quaternion is given by $\dot{q}=q\Omega/2$, where
the multiplication rule of quaternions is adopted \cite{Kuipers1999Quaternions,Coutsias2004The}, and $\Omega=(0,\Omega_{a},\Omega_{b},\Omega_{c})$. 
The ensemble average gives the degree of orientation,
\begin{align}
    \braket{\cos(\theta_{aj})}(t) = \frac{1}{N^\prime}\sum\limits_{n^\prime=1}^{N^\prime} \cos[\theta_{aj}(n^\prime,t)] ,
\end{align}
where $N^\prime=10^6$ is the number of sample molecules used in the classical ensemble and $\theta_{aj}(n^\prime,t)$ is the time-dependent angle of the $n^\prime$-th molecule.
A more detailed description of the classical simulations can be found in \cite{Xu2020longlasting}.

\section{Persistent orientation induced by a linearly polarized THz pulse}\label{sec:single_pulse}

We begin with the case of excitation by a single pulse. Consider asymmetric-top $\mathrm{CH_2O}$ molecules excited by a single THz pulse propagating along the laboratory $Z$ axis and polarized along the $X$ axis.
The electric field of the THz pulse is given by
\begin{align}
    \mathbf{E}(t) = E_0f(t)\mathbf{e}_{X},
    \label{eq:single_THz_Field}
\end{align}
where the time dependence is described by \cite{Coudert2017Optimal}
\begin{align}
    f(t) = (1-2\sigma t^2) e^{-\sigma t^2}, 
    \label{eq:envelope}
\end{align}
$E_0$ is the peak amplitude, $\sigma$ determines the temporal width of the THz pulse (in our simulations we use  $\sigma = 3.06\,\mathrm{ps}^{-2}$), and $\mathbf{e}_{X}$ is a unit vector along the laboratory $X$ axis.
\begin{figure}[!t]
\centering\includegraphics[width=\linewidth]{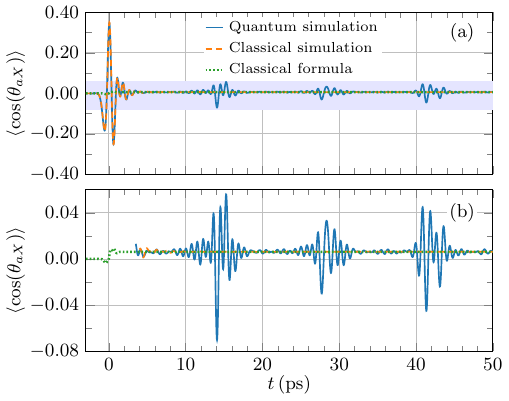}
    \caption{
    (a) Classically and quantum mechanically calculated time-dependent degree of orientation of $\mathrm{CH_2O}$ 
        molecules excited by a linearly polarized THz pulse. Here, $T=300\,\mathrm{K}$ 
        and the field amplitude is $E_0=6.5\,\mathrm{MV/cm}$ [see Eq. \eqref{eq:single_THz_Field}]. 
    (b) Magnified view of the rectangular region marked in (a). The value obtained using the        
        classical formula [$\braket{L_a L_X/L^2}$, see Eq. \eqref{eq:classical_formula}] is shown by the dotted green line. (see text)
           }
    \label{fig:fig1}
\end{figure}

Figure \ref{fig:fig1} shows the time-dependent degree of orientation, $\braket{\cos(\theta_{aX})}$. 
The rotational temperature is set to $T = 300\,\mathrm{K}$.
As expected, a strong transient orientation appears in response to the THz pulse. 
The results of quantum and classical simulations are in excellent agreement on the short time scale [see Fig. \ref{fig:fig1}(a)]. On a longer time scale, the classical orientation factor tends to a nonzero constant value [see Fig. \ref{fig:fig1}(b)], which we denote by $\braket{\cos(\theta_{aX})}_p$, while the quantum expectation value oscillates around the same nonzero baseline. 
This effect induced by a single THz pulse has been previously studied \cite{Xu2020longlasting}.
The long-lasting orientation metric, $\braket{\cos(\theta_{aX})}_p$ was termed the \emph{persistent orientation} factor. 

Next, we derive an approximate analytic expression for the persistent orientation factor in the case of asymmetric-top molecules.
Starting from the wave function of an asymmetric-top molecule after the THz excitation, at $t\,(t>t_f)$ 
\begin{align}\label{eq:psit}
    \ket{\Psi(t)}
    &= e^{-\frac{i}{\hbar}\int_{t_f}^t H_r dt^\prime}\ket{\Psi(t_f)} \nonumber\\
    & = \sum\limits_{J\kappa M} C_{J\kappa M} \ket{J\kappa M} e^{-\frac{i}{\hbar} \varepsilon_{J\kappa M}(t-t_f)},
\end{align}
where $t_f$ is a moment at the end of the pulse, $C_{J\kappa M} = \braket{J\kappa M|\Psi(t_f)}$,  and 
\begin{align}\label{eq:psitf}
\ket{\Psi(t_f)}= \mathcal{T}e^{-\frac{i}{\hbar}\int_{t_i}^{t_f} H(t^\prime) \, dt^\prime}\ket{\Psi(t_i)}
\end{align}
is the wave function immediately at the end of the pulse. $\mathcal{T}$ is the time-ordering operator.
Recall that the thermal effects are accounted for in the initial wave function with the random phases, as shown in Eq. \eqref{eq:initial_state}. Here, we use a single $|\Psi_n(t_i)\rangle$ which suffices to describe the dynamics at high enough temperatures. Accordingly, the time-dependent degree of orientation is given by
\begin{align}
   &\braket{\cos(\theta_{aX})}(t) = \braket{\Psi(t)|\cos(\theta_{aX})|\Psi(t)} \nonumber\\
   =&\sum\limits_{JKM} \sum\limits_{J^\prime, \kappa, \kappa^\prime} A_{JKM,\kappa}^* A_{J^\prime K M, \kappa^\prime}\braket{JKM|\cos(\theta_{aX})|J^\prime KM}\nonumber\\
   & \times e^{-\frac{i}{\hbar} (\varepsilon_{J^\prime \kappa^\prime M}-\varepsilon_{J\kappa M})(t-t_f)},
\end{align}
where $A_{JKM,\kappa}= \braket{JKM|J\kappa M}\braket{J\kappa M|\Psi(t_f)}$. 
The mean orientation factor approximates the persistent orientation factor, 
\begin{align}
    \braket{\cos(\theta_{aX})}_p 
    &\approx\frac{1}{T_\mathrm{rev}}\int_{t_f}^{t_f+T_\mathrm{rev}} \braket{\cos(\theta_{aX})}(t) dt \nonumber\\
    &\approx\sum\limits_{JKM} \sum\limits_{\kappa, \kappa^\prime} A_{JKM,\kappa}^* A_{JKM, \kappa^\prime}\frac{KM}{J(J+1)}, \!\!\!\label{eq:quantum_formula}
\end{align}
where $T_\mathrm{rev}$ is a ``revival period''. It should be noted that several revival times can be defined for asymmetric-top molecules. In the context of our derivation, $T_\mathrm{rev}$ refers to a typical rotational time scale on the order of tens of picoseconds, and variations in the exact value are insignificant.
Here, we neglect the oscillatory behavior caused by the nonzero energy difference between $|J \kappa M\rangle $ and $|J^\prime \kappa^\prime M\rangle$, $\Delta\varepsilon=\varepsilon_{J^\prime \kappa^\prime M}-\varepsilon_{J\kappa M}$.
Nonzero $\Delta\varepsilon$ produces oscillations, sign changes, and even the vanishing of the persistent orientation factor on the nanosecond time scale. Proper description of the molecular rotations on the nanosecond time scale may require considering intermolecular collisions, complicating the model considerably. Thus, we leave the study of persistent orientation on such a long time scale for future studies.

In the special case of a symmetric-top molecule, for which $K$ is a good quantum number, the formula in Eq. \eqref{eq:quantum_formula} reduces to (see Appendix \ref{sec:persistent-orientation} for details)
\begin{align}
    \braket{\cos(\theta_{aX})}_p &\approx\sum\limits_{JKM}  |A_{JKM,K}|^2
    \frac{KM}{J(J+1)} \nonumber\\
    &=\sum\limits_{JKM}  |\braket{JKM|\Psi(t_f)}|^2\frac{KM}{J(J+1)}, \label{eq:symmetric}
\end{align}
which is the same as in \cite{Liang2022Upper}. The corresponding classical formula, $\braket{L_aL_X/L^2}$, can be derived from a single classical symmetric top persistent orientation, $L_aL_X/L^2$ \cite{Xu2020longlasting,Xu2021longlasting}. Here, angle brackets denote the ensemble average, $L$ is the magnitude of the angular momentum, $L_a$ and $L_X$ are the angular momentum projections on the molecular $a$ axis, and the laboratory $X$ axis (polarization direction of the THz pulse), respectively. The direct relation between the quantum and classical formulas is evident due to the correspondence between the quantum and classical quantities, $J\leftrightarrow L$, $K\leftrightarrow L_a$, and $M\leftrightarrow L_X$. After the pulse, $L$, $L_a$, and $L_X$ are conserved in the case of a symmetric-top molecule, while $L_a$ and $L_X$ are conserved even during the pulse. 

The case of asymmetric-top molecules is more complicated because $L_a$ of the individual molecules is not a constant of the motion \cite{Goldstein2002Classical}. Nevertheless, we found that after the pulse, the distribution of $L_a L_X/L^2$ reaches a dynamic equilibrium, such that the ensemble average $\braket{L_a L_X/L^2}$ is effectively time-independent (like in the symmetric-top case). Thus, the classical formula, corresponding to the quantum-mechanical one in Eq. \eqref{eq:quantum_formula} is
\begin{align}
    \braket{\cos(\theta_{aX})}_p \approx
    \left\langle \frac{L_a L_X}{L^2}\right\rangle
    =\frac{1}{N^\prime}\sum\limits_{n^\prime=1}^{N^\prime} \frac{L_a (n^\prime,t) L_X(n^\prime)}{L^2(n^\prime)},
    \label{eq:classical_formula}
\end{align}
where $L(n^\prime)$, $L_a(n^\prime,t)$, and $L_X(n^\prime)$ are the angular momentum magnitude and components of the $n^\prime$-th molecule. 
After the pulse, $L_a(n^\prime,t)$ is time-dependent, while $L_X(n^\prime)$ and  $L(n^\prime)$ are conserved.
Figure \ref{fig:fig1}(b) compares the prediction of the formula in Eq. \eqref{eq:classical_formula} with the results of fully time-dependent simulation.
Equation \eqref{eq:classical_formula} shows an excellent approximation to the persistent orientation factor and reflects the following qualitative symmetry argument. If the molecules are initially isotropically distributed in space, the distribution of $L_a L_X/L^2$ is fully symmetric, which implies that $\braket{L_a L_X/L^2}=0$. The THz pulse(s) breaks the symmetry of the molecular ensemble, resulting in nonzero average $\braket{L_a L_X/L^2} \neq 0$ during and after the pulse \cite{Xu2020longlasting}.

\begin{figure}[!t]
\centering{}
\includegraphics[width=\linewidth]{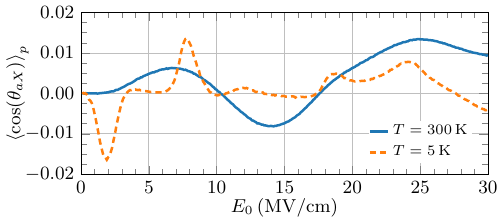}
    \caption{
            Classically calculated persistent orientation factor as a function of field amplitude at low ($5\,\mathrm{K}$) and at room ($300\,\mathrm{K}$) rotational temperatures. 
            }
    \label{fig:fig2}
\end{figure}

It is instructive to consider the dependence of the persistent orientation factor, $\braket{\cos(\theta_{aX})}_p$ on the peak amplitude of the THz pulse, $E_0$ [see Eq. \eqref{eq:single_THz_Field}]. Figure \ref{fig:fig2} shows the classically calculated persistent orientation factor as a function of the field amplitude, $\braket{\cos(\theta_{aX})}_p(E_0)$. The degree of orientation is taken at $t=50\,\mathrm{ps}$, after the steady-state value has been reached, see Fig. \ref{fig:fig1}(b).
For linearly polarized THz fields, with increasing field amplitudes, the value of the persistent orientation factor fluctuates between positive and negative values, and when $E_0\leq 30\,\mathrm{MV/cm}$, its absolute value is limited: $|\braket{\cos (\theta_{aX})}_p|\leq 0.013$ at $T=300\,\mathrm{K}$ and $|\braket{\cos (\theta_{aX})}_p|\leq 0.016$ at $T=5\,\mathrm{K}$. The persistent orientation factor, $\braket{\cos(\theta_{aX})}_p(E_0)$, is a non-monotonic function at both low and high temperatures, which is consistent with our previous study \cite{Xu2020longlasting}. Thus, increasing the field amplitude won't necessarily lead to a higher value of the persistent orientation.

For field polarization fixed along the $X$ axis, $L_X$ remains constant during and after the pulse. Thus, considering the formula in Eq. \eqref{eq:classical_formula}, when $E_0 \rightarrow 0$, the symmetry of the distribution of $L_a L_X/L^2$ remains mostly unchanged, resulting in vanishing persistent orientation.
When $E_0 \rightarrow \infty$, the ensemble symmetry may be completely broken, but the post-pulse angular momentum $L$ also tends to infinity, resulting in $L_a L_X/L^2 \rightarrow 0$. Thus, between these limits, there will be a value of the field amplitude that will give rise to the maximal persistent orientation.

One way to potentially enhance the persistent orientation factor is by shaping the \emph{field polarization}. The qualitative arguments above do not apply in this case because the angular momentum projection $L_X$ is no longer conserved when the field polarization is not fixed (e.g., a pair of time-delayed cross-polarized THz pulses). The following section explores this potential route to enhanced persistent orientation.

\section{Persistent orientation induced by a pair of time-delayed cross-polarized THz pulses}\label{sec:delayed_pulse}

\begin{figure*}[!t]
\centering{}
\includegraphics[width=15cm]{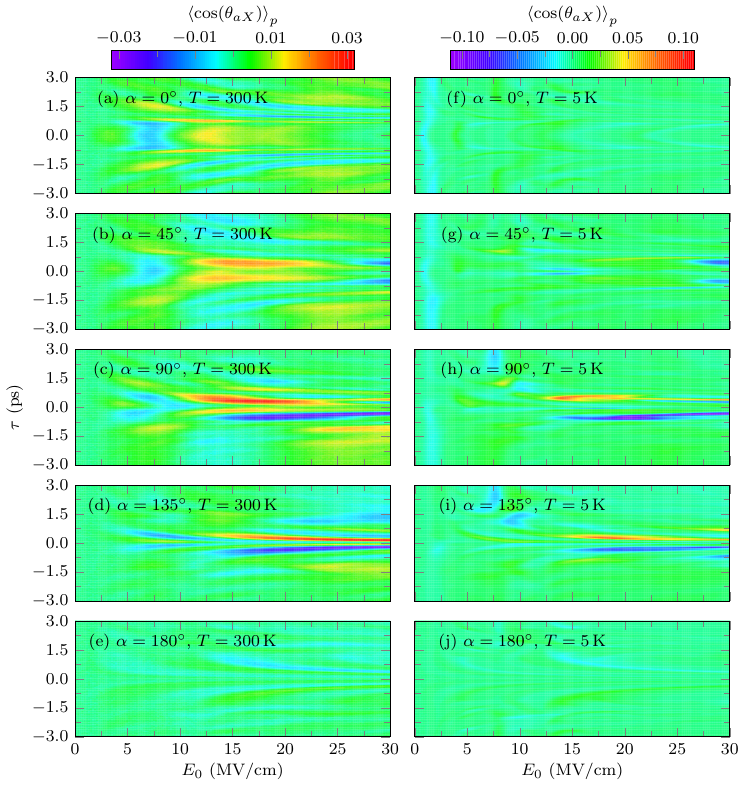}
\caption{
        Classical persistent orientation factor, $\braket{\cos(\theta_{aX})}_p$ as a function of field 
        amplitude, $E_0$ and time delay, $\tau$ for several relative angles, $\alpha$ [see Eqs. \eqref{eq:two_THz_Field} and \eqref{eq:polarization_angle}] 
        at $T=300\,\mathrm{K}$ (left panels) and $T=5\,\mathrm{K}$ (right panels).
        As expected, the (absolute) values are lower at room temperature (note the difference in color bar scales).
        The persistent orientation factor along $\mathbf{e}_2$ (the polarization direction of the second THz pulse) satisfies $\braket{\cos(\theta_{a\mathbf{e}_2})}_p(E_0,-\tau)=\braket{\cos(\theta_{aX})}_p(E_0,\tau)$.
        }
        \label{fig:fig3}
\end{figure*}

\begin{figure*}[!t]
\centering{}
\includegraphics[width=15cm]{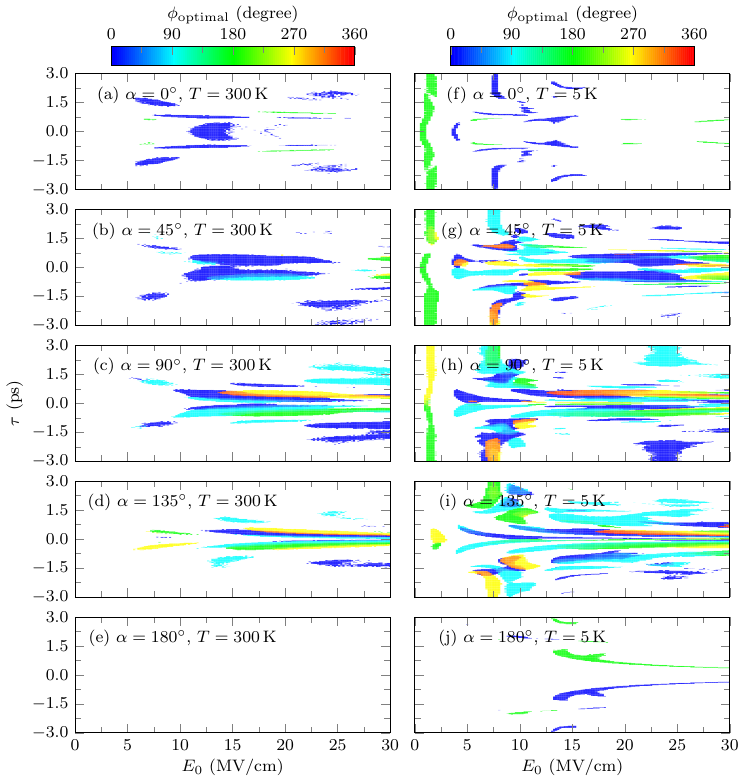}
\caption{
        Classically calculated $\phi_{\mathrm{optimal}}$ (along which the maximum persistent orientation factor is reached) as a function of field amplitude and time delay for different angles $\alpha$ [see Eqs. \eqref{eq:two_THz_Field} and \eqref{eq:polarization_angle}] 
        at $T=300\,\mathrm{K}$ (left panels) and $T=5\,\mathrm{K}$ (right panels). Only persistent orientation factor values larger than 0.01 are considered.
        The field parameters used are the same as in Fig. \ref{fig:fig3}.
        }
        \label{fig:fig41}
\end{figure*}

\begin{figure*}[!t]
\centering{}
\includegraphics[width=15cm]{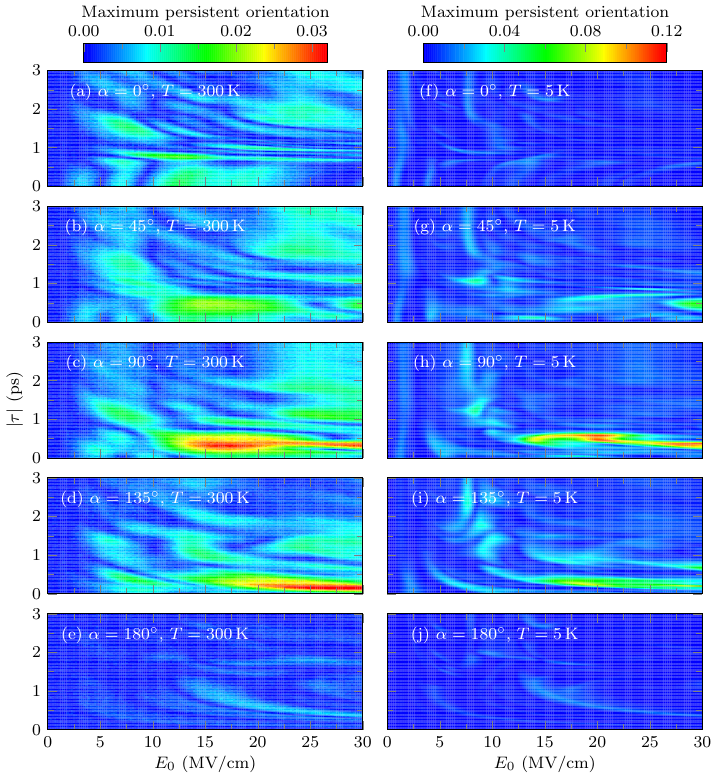}
    \caption{
            Classically calculated maximum persistent orientation factor in the $XY$ plane as a function of field amplitude 
            and time delay (in absolute value) for various angles $\alpha$ [see Eqs. \eqref{eq:two_THz_Field} and \eqref{eq:polarization_angle}] 
            at $T=300\,\mathrm{K}$ (left panels) and $T=5\,\mathrm{K}$ (right panels).
            The field parameters used are the same as in Fig. \ref{fig:fig3}.
            Equal maximum persistent orientation factors are achievable for positive or negative time delays but along different $\phi$.
            }
    \label{fig:fig5}
\end{figure*}

Consider two pulses propagating along the laboratory $Z$ axis; the first THz pulse is polarized along the $X$ direction, while the second one is polarized in the $XY$ plane at angle $\alpha$ to the $X$ direction.
The resulting electric field is described by
\begin{equation}
\mathbf{E}(t)=E_0 \Big[f(t)\mathbf{e}_{X}+f(t-\tau)\mathbf{e}_2\Big],\label{eq:two_THz_Field}
\end{equation}
where the polarization direction of the second THz pulse is given by
\begin{equation}
\mathbf{e}_2=\cos(\alpha)\mathbf{e}_{X} + \sin(\alpha)\mathbf{e}_{Y},\label{eq:polarization_angle}
\end{equation}
$\mathbf{e}_{Y}$ is a unit vector along the laboratory $Y$ axis, $f(t)$ is defined in Eq. \eqref{eq:envelope}, and $\tau$ is the time delay between the peaks of two THz pulses.

 We wish to investigate the dependence of the persistent orientation on the various parameters: field amplitude $E_0$,  the delay between the two pulses $\tau$, and the angle between their individual polarizations $\alpha$.  Figure \ref{fig:fig3} depicts this dependence for two rotational temperatures: $T=5\,\mathrm{K}$ (right side) and  room ($T=300\,\mathrm{K}$) (left side). For each temperature we show five panels, each for a different angle $\alpha$ between the two polarizations: $\alpha=0^\circ, 45^\circ, 90^\circ, 135^\circ, 180^\circ $. Each panel shows the molecular persistent orientation factor along the laboratory $X$ axis as a function of the field amplitude (horizontal axis) and the time delay (vertical axis), $\braket{\cos(\theta_{aX})}_p(E_0,\tau)$. The value of the persistent orientation is given by the color code.

A sizable persistent orientation factor appears at short delays, $0.2\,\mathrm{ps}<|\tau|<0.6\,\mathrm{ps}$, when the two time-delayed THz pulses partially overlap. Note, however, for $\tau \approx 0$, the two THz pulses coalesce into a single linearly polarized pulse, resulting in a small persistent orientation factor.

By observation, under some conditions, the cross-polarized pulses induce a higher persistent orientation factor (colors blue or red in the figure). At $\alpha=45^\circ, 135^\circ$ and especially for $\alpha=90^\circ$, the values are much higher than for the collinearly polarized pulses ($\alpha=0^\circ, 180^\circ$). The most efficient configuration is for  $\alpha=90^\circ$. Indeed, the optimal polarization angle, $\alpha$, depends on the temperature, the field amplitude, and the time delay.
Figure \ref{fig:fig3}(c) shows that at $T=300\,\mathrm{K}$, the highest (in absolute value) persistent orientation factor, $\braket{\cos (\theta_{aX})}_p\approx 0.029$, is achieved when $E_0=17\,\mathrm{MV/cm}$ and $\tau=0.32\,\mathrm{ps}$. 
Note that a slightly higher value ($\braket{\cos (\theta_{aX})}_p\approx 0.032$) is achieved in the case of $\alpha=135^\circ$ [see Fig. \ref{fig:fig3}(d)], but it requires a much higher field amplitude ($E_0=30\,\mathrm{MV/cm}$).

The use of cross-polarized pulses at room temperature almost doubles the persistent orientation induced by a pair of collinearly polarized pulses [$|\braket{\cos (\theta_{aX})}_p|< 0.019$ in Fig. \ref{fig:fig3}(a) and $|\braket{\cos (\theta_{aX})}_p|< 0.008$ in Fig. \ref{fig:fig3}(e)].
The enhancement is more pronounced at low temperatures.
As shown in Fig. \ref{fig:fig3}(h) for $T=5\,\mathrm{K}$, a value of $\braket{\cos (\theta_{aX})}_p\approx -0.101$ is reached when $E_0=21\,\mathrm{MV/cm}$ and $\tau=-0.58\,\mathrm{ps}$. This is more than five times higher (in absolute value) compared to collinearly polarized THz pulses [$|\braket{\cos (\theta_{aX})}_p|< 0.019$ in Figs. \ref{fig:fig3}(f) and \ref{fig:fig3}(j)].
These findings confirm that polarization shaping (e.g., combining two time-delayed orthogonally polarized pulses) may be efficient for increasing the persistent orientation. 

A single THz pulse induces persistent orientation along its polarization direction. However, cross-polarized THz pulses induce orientation in the $XY$ plane, and it is not apriori clear that the polarization along the $X$ direction is the maximal one. For fixed field parameters (polarization angle, field amplitude, and time delay), we consider the persistent orientation along a direction in the $XY$ plane defined by 
$\mathbf{e}_{XY}=\cos(\phi)\mathbf{e}_{X} + \sin(\phi)\mathbf{e}_{Y}$, where the angle $\phi$ runs from $0^\circ$ to $360^\circ$.
The direction along which the maximum persistent orientation factor appears is defined as  $\phi_{\mathrm{optimal}}$.
We plot $\phi_{\mathrm{optimal}}$ and maximum persistent orientation factor (measured along the optimal direction) as functions of the field amplitude and the time delay, for each set of field parameters in Figs. \ref{fig:fig41} and \ref{fig:fig5}, respectively.
For better visibility, Fig. \ref{fig:fig41} shows only the points where the maximum persistent orientation factor is larger than 0.01.
As can be seen, $\phi_{\mathrm{optimal}}$ has a non-monotonic dependence on the field amplitude and the time delay.

Figure \ref{fig:fig5} shows that sizable persistent orientation factor appears when $0.2\,\mathrm{ps}<|\tau|<0.6\,\mathrm{ps}$, similar to $\braket{\cos (\theta_{aX})}_p$ in Fig. \ref{fig:fig3}.
For the collinearly polarized THz pulses ($\alpha=0^\circ, 180^\circ$), the maximum persistent orientation factor is the same as in Fig. \ref{fig:fig3}. However, the maximum persistent orientation factor is slightly higher in the case of cross-polarized THz pulses.
Figure \ref{fig:fig5}(c) shows that at $T=300\,\mathrm{K}$, the global maximum at $E_0=17\,\mathrm{MV/cm}$ and $\tau=0.32\,\mathrm{ps}$ is about 0.0293, which is a slightly higher than $\braket{\cos (\theta_{aX})}_p$ in Fig. \ref{fig:fig3}.
At $T=5\,\mathrm{K}$, the global maximum is about 0.116 at $E_0=30\,\mathrm{MV/cm}$ and $\tau=0.38\,\mathrm{ps}$, as shown in Fig. \ref{fig:fig5}(h).
\begin{figure}[!t]
\centering{}
\includegraphics[width=\linewidth]{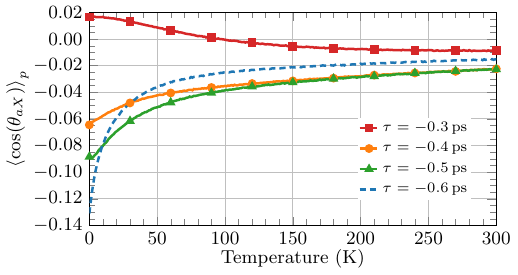}
\caption{Classically calculated persistent orientation factor as a function of temperature for different time delays. Here the field amplitude is fixed at $E_0=21\,\mathrm{MV/cm}$ and the polarization angle is fixed at $\alpha=90^\circ$, see Eqs. \eqref{eq:two_THz_Field} and \eqref{eq:polarization_angle}.
 \label{fig:fig4}}
\end{figure}

Temperature is a crucial parameter affecting the persistent orientation \cite{Tutunnikov2019Laser,Xu2020longlasting,Tutunnikov2021Enantioselective,Xu2021longlasting,Xu2022Ionization}.
Figure \ref{fig:fig4} shows the temperature dependence of the persistent orientation factor, $\braket{\cos(\theta_{aX})}_p$ for different time delays.
Here, we consider orthogonally polarized THz pulses ($\alpha=90^\circ$) and the field amplitude is fixed at $E_0=21\,\mathrm{MV/cm}$.
Recall that the persistent orientation factor along the laboratory $X$ axis is approximately given by $\braket{L_a L_X/L^2}$. 
In the case of a single $X$-polarized THz pulse, there is no persistent orientation at $T=0\,\mathrm{K}$ because the conserved $L_X$ is zero \cite{Xu2020longlasting}.
In contrast, the orthogonally polarized pulses change $L_X$ and can induce nonzero persistent orientation factor in the $XY$ plane even at $T=0\,\mathrm{K}$.

For long time delays,  the maximum (in absolute value) $\braket{\cos(\theta_{aX})}_p\approx -0.131$ appears at $T=0\,\mathrm{K}$, and $\braket{\cos(\theta_{aX})}_p$ decays with the temperature (see the case of $\tau=-0.6\,\mathrm{ps}$ in Fig. \ref{fig:fig4}).
The persistent orientation factor slowly decreases with temperature for shorter time delays (see the cases of $\tau=-0.5,-0.4\,\mathrm{ps}$).
The optimal time delay at higher temperatures is shorter than at low temperatures.
For instance, at $T = 0\,\mathrm{K}$, the optimal time delay is $\tau=-0.6\,\mathrm{ps}$, while at $T = 300\,\mathrm{K}$, the optimal time delay is $\tau=-0.4\,\mathrm{ps}$.
The effective rotational time scale is shorter at high temperatures, requiring shorter time delays.
As noted, for vanishingly short time delays, the two pulses coalesce to a linearly polarized pulse, leading to a smaller persistent orientation factor that may even change its sign with temperature (see the case of $\tau=-0.3\,\mathrm{ps}$).

\section{Conclusions}\label{sec:conclusions}
 
We set out to investigate the persistent orientation of asymmetric-top molecules excited by a pair of cross-polarized THz pulses, aiming to overcome the limitations of the existing scheme that yields relatively small persistent orientation.
Our simulations revealed that using time-delayed, orthogonally polarized THz pulses significantly enhances the persistent orientation, especially at lower temperatures. These findings not only outperform previous methods but also suggest that measurements at room temperature are now within experimental feasibility.

Furthermore, we derived an approximate expression describing the field-free persistent orientation of asymmetric-top molecules. 
One key distinction of our cross-polarized scheme is that it disrupts the conservation of angular momentum along the field polarization direction, contributing to the observed enhancement. While this qualitative insight provides an initial explanation, a more detailed and rigorous understanding of the underlying mechanism still warrants further research. Moreover, simultaneous optimization of all controllable parameters (including, for example, pulse duration) may yield even greater improvements.

The persistent orientation may be measured by probing the free-induction-decay \cite{Fleischer2011Molecular,Babilotte2016Observation}, by measuring the second (or higher order) harmonic generation in the gas phase \cite{Frumker2012Oriented,Frumker2012Probing,Kraus2012High}, or via Coulomb explosion \cite{Znakovskaya2014Transition,Egodapitiya2014Terahertz,Tutunnikov2020Observation}. Naturally, intermolecular collisions will eventually destroy the persistent orientation. However, this may be useful, as time-resolved measurement of persistent orientation decay may provide new insights into the molecular gas relaxation dynamics.

\begin{acknowledgments}
This work is supported by the
startup funding from Xiamen University. I.A. gratefully acknowledges the hospitality extended to him during his stay at the Department of Chemistry of the University of British Columbia. This research was made possible in part by the historic generosity of the Harold Perlman Family.

\end{acknowledgments}
\appendix
\section{Persistent orientation factor of asymmetric-top molecules}\label{sec:persistent-orientation}

The time-dependent degree of orientation along the laboratory $X$ axis, according to Eq. \eqref{eq:psit}, is given by 
\begin{widetext}
\begin{align}
  & \braket{\cos(\theta_{aX})}(t) = \braket{\Psi(t)|\cos(\theta_{aX})|\Psi(t)} \nonumber\\
   =&\sum\limits_{J\kappa M} \sum\limits_{J^\prime \kappa^\prime M^\prime } C_{J\kappa M}^* C_{J^\prime \kappa^\prime M^\prime }\braket{J\kappa M|\cos(\theta_{aX})|J^\prime  \kappa^\prime M^\prime} e^{-\frac{i}{\hbar} (\varepsilon_{J^\prime \kappa^\prime M^\prime}-\varepsilon_{J\kappa M})(t-t_f)}\nonumber\\
   =&\sum\limits_{J\kappa M} \sum\limits_{J^\prime \kappa^\prime M^\prime}\sum\limits_{K,K^\prime} C_{J\kappa M}^* C_{J^\prime  \kappa^\prime M^\prime}\braket{J\kappa M|JKM}\braket{JKM|\cos(\theta_{aX})|J^\prime  K^\prime M^\prime}\braket{J^\prime K^\prime M^\prime|J^\prime  \kappa^\prime M^\prime} e^{-\frac{i}{\hbar} (\varepsilon_{J^\prime  \kappa^\prime M^\prime}-\varepsilon_{J\kappa M})(t-t_f)}\nonumber\\
   =&\sum\limits_{JKM} \sum\limits_{J^\prime,\kappa, \kappa^\prime} C_{J\kappa M}^* C_{J^\prime\kappa^\prime  M}\braket{J\kappa M|JKM}\braket{JKM|\cos(\theta_{aX})|J^\prime K M}\braket{J^\prime K M|J^\prime \kappa^\prime M} e^{-\frac{i}{\hbar} (\varepsilon_{J^\prime \kappa^\prime M}-\varepsilon_{J\kappa M})(t-t_f)}\nonumber\\
   =&\sum\limits_{JKM} \sum\limits_{J^\prime, \kappa, \kappa^\prime} A_{JKM,\kappa}^* A_{J^\prime K M, \kappa^\prime}\braket{JKM|\cos(\theta_{aX})|J^\prime K M}e^{-\frac{i}{\hbar} (\varepsilon_{J^\prime \kappa^\prime M}-\varepsilon_{J\kappa M})(t-t_f)},
    \label{eq:td_orientation}
\end{align}
\end{widetext}
where $A_{JKM,\kappa}=C_{J\kappa M}\braket{JKM|J\kappa M}=\braket{JKM|J\kappa M}\braket{J\kappa M|\Psi(t_f)}$.

The persistent orientation factor is approximated as the mean orientation factor averaging over a revival period, 
\begin{align}
    \braket{\cos(\theta_{aX})}_p 
    &\approx\frac{1}{T_\mathrm{rev}}\int_{t_f}^{t_f+T_\mathrm{rev}} \braket{\cos(\theta_{aX})}(t) dt \nonumber\\
   &\approx\sum\limits_{JKM} \sum\limits_{\kappa, \kappa^\prime} A_{JKM,\kappa}^* A_{JKM, \kappa^\prime}\frac{KM}{J(J+1)}, \!\!\!\label{eq:asymmetric}
\end{align}
where $T_\mathrm{rev}$ is a ``revival period''. It should be noted that several revival times can be defined for asymmetric-top molecules. In the context of our derivation, $T_\mathrm{rev}$ refers to a typical rotational time scale on the order of tens of picoseconds, and variations in the exact value are insignificant.
Here, we neglect the oscillatory behavior caused by the coherence between $|J \kappa M\rangle $ and $|J^\prime \kappa^\prime M\rangle$, which has a nonzero energy difference, $\Delta\varepsilon=\varepsilon_{J^\prime \kappa^\prime M}-\varepsilon_{J\kappa M}$.
On the long-time scale, such nonzero energy difference may result in oscillation, sign change, and even the vanishing of the persistent orientation factor.
This is in sharp contrast to the case of symmetric-top molecules, where the persistent orientation is permanent.

The states $|JKM\rangle$ are the eigenfunctions of the symmetric-top molecules, thus the time-dependent degree of orientation [see Eq. \eqref{eq:td_orientation}] can be expressed as
\begin{align}
  & \braket{\cos(\theta_{aX})}(t)  \nonumber\\
  =&\sum\limits_{J^\prime,JKM}  A_{JKM,K}^* A_{J^\prime K M, K}\braket{JKM|\cos(\theta_{aX})|J^\prime K M}\nonumber\\
   & \times e^{-\frac{i}{\hbar} (\varepsilon_{J^\prime K M}-\varepsilon_{JK M})(t-t_f)}.
  \label{eq:td_orientation_S}
\end{align}
For symmetric-top molecules [$B=C$, see Eq. \eqref{eq:HR}], the rotational kinetic energy takes the form \cite{zare1988Angular}
\begin{align}
\begin{split}
\varepsilon_{JKM}=B J(J+1)+(A-B)K^{2}.
\end{split}
\end{align}
As a result, the energy difference between $|JKM\rangle $ and $|J^\prime KM\rangle $ is $\varepsilon_{J^\prime KM}-\varepsilon_{JKM}=B(J^\prime+J+1)(J^\prime-J)=2mB$, where $m$ is an integer.
Averaging over a revival period, $T_\mathrm{rev}=\pi \hbar/B$ \cite{Robinett2004}, the accumulated phase is $-(\varepsilon_{J^\prime KM}-\varepsilon_{JKM})T_\mathrm{rev}/\hbar=-2m\pi$, resulting in the persistent orientation factor
\begin{align}
    \braket{\cos(\theta_{aX})}_p &\approx\sum\limits_{JKM}  A_{JKM,K}^* A_{JKM,K}\frac{KM}{J(J+1)} \nonumber\\
   &=\sum\limits_{JKM}  |\braket{JKM|\Psi(t_f)}|^2\frac{KM}{J(J+1)}, \label{eq:symmetric}
\end{align}
which is consistent with the results in the quantum \cite{Liang2022Upper} and classical cases \cite{Xu2020longlasting,Xu2021longlasting}.

The persistent orientation factors in the cases of symmetric- [Eq. \eqref{eq:symmetric}] and asymmetric-top [Eq. \eqref{eq:asymmetric}] molecules are similar, and both depend on the three quantum numbers: $J$, $K$, $M$.
However, there are two main differences:
(i) The persistent orientation factor is permanent for symmetric-top molecules, while it oscillates and eventually vanishes for asymmetric-top molecules due to the nonzero energy difference, $\Delta\varepsilon=\varepsilon_{J^\prime \kappa^\prime M}-\varepsilon_{J\kappa M}$.
(ii) According to Eqs. \eqref{eq:asymmetric} and \eqref{eq:symmetric}, the persistent orientation factor depends on the quantum number $K$. As a result, each symmetric-top molecule has a specific contribution to the persistent orientation factor. On the other hand, in the case of an asymmetric-top molecule, $K$ is not a good quantum number; the contribution of each molecule is time-dependent and only the ensemble has a well-defined, approximately time-independent persistent orientation factor.

\end{document}